\documentclass[aps,prl,onecolumn,superscriptaddress,groupedaddress]{revtex4}  
\usepackage{graphicx}  
\usepackage{dcolumn}   
\usepackage{bm}        
\usepackage{amssymb}   
\usepackage{amsmath}
\usepackage{xcolor}
\begin{document}

\title{Sequence Effects on Internal Structure of Droplets of Associative Polymers}

\author{ Kulveer Singh }
\email{kulveersingh85@gmail.com}
\author{Yitzhak Rabin}%
 \email{yitzhak.rabin@biu.ac.il}
\affiliation{Department of Physics, and Institute of Nanotechnology and Advanced Materials, \\Bar-Ilan University,
Ramat Gan 52900, Israel}%

\date{\today}

\begin{abstract}
We used Langevin dynamics simulations of short associative polymers with two stickers placed symmetrically along their contour to study the effect of the primary sequence of these polymers on their organization inside condensed droplets. We observed that the shape, size and number of sticker clusters inside the condensed droplet change from a single cylindrical fiber to many compact clusters, as one varies the location of stickers along the chain contour. Aging due to conversion of intramoleclular to intermolecular associations was observed in droplets of telechelic polymers, but not for other sequences of associating polymers.
The relevance of our results to condensates of intrinsically disordered proteins is discussed. 
\end{abstract}

\maketitle



\section{Introduction}

Membrane-less subcellular compartments such as P granules, nucleoli, cajal bodies, etc., perform specialized biochemical roles inside cells \cite{Brangwynne2009, Brangwynne2011}. The formation of these biomolecular condensates is governed by liquid-liquid phase separation where biopolymers such as proteins and nucleic acids condense into liquid droplets \cite{Hyman2014, Brangwynne2017, Hyman2017, Alberti2019}. This phase separation depends on various factors such as polymer-polymer, polymer-solvent and solvent-solvent interactions, concentration of polymers in solvent and environmental conditions such as temperature, pH, etc.  A significant fraction of all proteins in a cell are flexible proteins which do not adopt a well-defined three dimensional structure and are known as intrinsically disordered proteins (IDPs) \cite{Oldfield2005}. Studies have revealed that IDPs are important ingredients of most biomolecular condensates in cells\cite{Wei2017, Protter2018, Majumdar2019}. A characteristic feature of IDPs is that their backbone contains short sequences of hydrophobic aminoacids that are strung together by flexible linkers that consist of  hydrophilic aminoacids \cite{Dyson2005}. These hydrophobic segments facilitate phase separation and gelation of IDPs in solution and give rise to variety of self-assembled structures such as micelles \cite{Klass2019}. 

Because of the presence of strongly associating sequences, IDPs can be considered as biological equivalents of associative polymers which contain segments or blocks of monomers known as stickers, that promote aggregation of these polymers in selective solvents \cite{Chassenieux2011, Colby}. Associative polymers undergo gelation (formation of system-spanning polymer networks) by forming physical crosslinks between stickers at sufficiently high concentration \cite{Semenov1, Semenov2, Rubinstein1999, Dobrynin2004, Dino2018}, and form flower-like micelles at low concentration \cite{Borisov1995}.  One example of such associative polymers are telechelic polymers which contain stickers at the two ends of the polymer chain. In aqueous solution telechelic polymers with hydrophobic stickers form flowerlike micelles which are connected (bridged) by other telechelic polymers  with two ends in two different micelles \cite{Semenov1995, Kulveer2020}. Upon their formation gels made of associative polymers crosslinked by clusters of stickers show aging behavior as they relax towards equilibrium \cite{Rabin2008, Gomez2013, Piazza2013},  due to slow structural reorganization produced by interconversion of intermolecular and  intramolecular assocations between stickers \cite{Kulveer2020}.

If the average inter-polymer attraction exceeds solvent-solvent and polymer-solvent interactions (poor solvent conditions), a solution of these associative polymers/IDPs undergoes phase separation into a polymer-rich phase that coexists with a dilute polymer solution \cite{Colby, Brangwynne2015}. If the average polymer concentration is sufficiently small, the process will take place via formation of droplets of the polymer-rich minority phase, that will grow by polymer exchange and by coalescence of droplets \cite{Lifshitz1961, Binder1977}. While this process has much in common with phase separation of homogeneous (i.e., made of identical monomers) polymers, the presence of strong associations between the stickers raises interesting questions about the internal morphology of these droplets. In particular, one would like to  characterize the size and the shape of clusters of stickers inside the droplets and to establish the connection between the internal morphology of the droplets and the sequence (primary structure) of the associative polymers/IDPs. One would also like to explore the kinetics of droplet formation and the temporal evolution of its internal structure. Finally, one would  like to understand what happens on the molecular level i.e.,  whether and how the balance between interchain and intrachain associations changes with time following the onset of phase separation. 

In order to address these questions, in Model and Methods section we introduce a simple model of associating polymers having two stickers symmetrically positioned along their contour. In Results section we use Langevin dynamics to simulate the relaxation of a dilute associating polymer solution to equilibrium, following a fast quench (e.g., by change of temperature \cite{Rabin2008} or pH \cite{Lauber2017} or  by rapid mixing in a microfluidic device \cite{Lemke2020}) to poor solvent conditions. We study the evolution of internal structure of large droplets (morphology of clusters of stickers) and the kinetics of interconversion between intramolecular and intermolecular associations, for different sequences of our model polymers. In Discussion secton we summarize our results on the polymer sequence dependence of the internal morphology and of the observed aging phenomena and discuss possible ramifications of our results for experiments on liquid IDP droplets.

\section{Model and Methods} 
We have performed implicit-solvent simulations of a solution of $M$ polymers of N=10 monomers (beads). Beads interact with each other via Lennard-Jones (LJ) potential given by  
\begin{equation}
	U^{LJ}_{ij}(r)=4\epsilon_{ij}\left[\left(\frac{\sigma}{r}\right)^{12}-\left(\frac{\sigma}{r}\right)^{6}\right]
\label{eq:LJ}
\end{equation}
which is truncated and shifted to zero at cutoff distance $r^{cut}_{ij}$ such that
\begin{equation}
U_{ij}^{LJ}(r) =
  \begin{cases}
	  U_{ij}^{LJ}(r) - U_{ij}^{LJ}(r_{ij}^{cut}) & r \le r_{ij}^{cut} \\	 0 & r > r_{ij}^{cut}
  \end{cases}.
\end{equation} 
Each polymer contains two types of beads designated as stickers and non-stickers respectively, such that $\epsilon_{ij} = \epsilon_s$ if $i^{th}$ and $j^{th}$ beads are stickers and  $\epsilon_{ij} = \epsilon_{ns}$ if at least one of those beads is a non-sticker. Neighboring beads along the backbone of the chain interact via finitely extensible nonlinear elastic (FENE) potential given by
\begin{equation}
	U^{FENE}=-0.5KR_0^2ln \left[ 1-\left(\frac{r}{R_0}\right)^2 \right]
\label{eq:fene},
\end{equation}
where we take $K=30.0$ and $R_0=1.5$. We use LAMMPS \cite{Plimpton1995} (Large-scale Atomic/Molecular Massively Parallel Simulator) to carry out Langevin dynamics simulations in the NVT ensemble. The simulation is performed in a box of size $61\times 61\times 61$ in units of $\sigma$, using periodic boundary conditions. The  motion of each bead is given by the Langevin equation, neglecting hydrodynamic interactions
\begin{equation}
m\ddot{\bf{r}}_i(t) = -\frac{\partial U}{\partial {\bf{r}}_i} -\zeta \dot{\bf{r}}_i(t) + \eta_i(t)
\label{eq:langevin},
\end{equation}
where $U$ (sum over all $U_{ij}$),  $\zeta$ and $\eta_i$ are the total potential energy, bead friction coefficient and random thermal force due to implicit solvent, respectively.  The rms amplitude of the random noise is proportional to  $(\zeta k_BT/\Delta t)^{1/2}$, where $k_B$, $T$ and $\Delta t$ are Boltzmann's constant, temperature and integration time step, respectively.  In the following, all the time scales are expressed in LJ time units $\tau_{LJ} = (m\sigma^2/\epsilon)^{1/2}=1$ (mass $m$, particle diameter $\sigma$, interaction parameter $\epsilon$, and temperature $k_BT$ are all set to 1).
We set the integration time-step to be $\Delta t =0.005$ and friction coefficient $\zeta = 0.02$. We took two stickers per polymer which were symmetrically placed along it contour (see Fig. \ref{fgr:schematic}). To obtain an initially uniform polymer solution, we placed all the polymers in an array inside the simulation box and equilibrated the system under good solvent conditions. We then changed the interaction parameters to poor solvent conditions and continued to monitor the system through the processes of drop formation and aging.

\section{Results}
\begin{figure}[h]
\centering
  \includegraphics[width=0.4\linewidth]{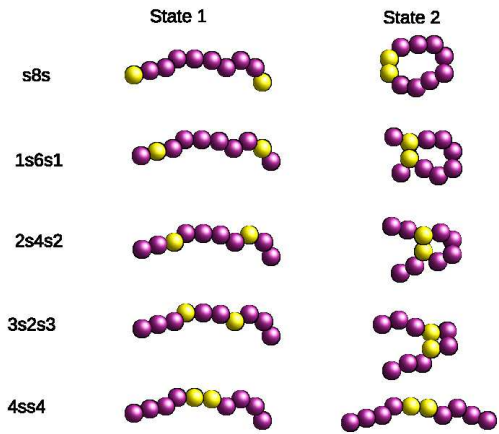}
  \caption{Open and bound states of 5 different sequences of  $N=10$ bead polymer, with two stickers symmetrically positioned along its contour.}
  \label{fgr:schematic}
\end{figure}

 We begin each simulation with dilute polymer solution in good solvent. The polymer volume fraction, $\phi = 0.011$, is chosen to be below the  overlap volume fraction $\phi^{\ast} \approx 0.27$ defined as the volume fraction of a single polymer in its pervaded volume. In order to ensure  good solvent conditions in the state of preparation we take $\epsilon_s=\epsilon_{ns}=0.8$ with cutoff distance $r^{cut}_{ij}=2^{1/6}\sigma$, corresponding to purely repulsive interactions between all beads. Simulations are then performed starting with a random initial configuration obtained by equilibrating the system under good solvent conditions. 

We first looked into the formation of a polymer droplet under poor solvent conditions assuming the same Lennard-Jones interaction between all beads, $\epsilon_s =\epsilon_{ns}=0.8$ and $r^{cut}_{ij}=2.5\sigma$ (note that this cutoff corresponds to both short range repulsion and long-range attraction between the beads). We verified that with this choice of parameters, phase separation between polymers and solvent occurs and a spherical polymer droplet condenses out of the solution (not shown). 

Since our aim is to study the effect of primary sequence of associative polymers on their organization inside condensed droplets, we model each polymer as a chain of eight weakly attractive beads and  two strongly attractive stickers and vary the location of the stickers along its contour.  The five different symmetric sequences of such a polymer shown in  Fig. \ref{fgr:schematic} range from the $s8s$ sequence which has two stickers at the ends and eight non-sticker beads in between (a telechelic polymer), to the $4ss4$ sequence in which the two stickers at the center of the chain  are flanked by four bead long tails (here $s$ denotes a sticker and the number specifies the length of a sequence of non-sticker beads).  The two possible states  of the polymers are depicted in Fig. \ref{fgr:schematic} where states 1 and 2 represent open and closed loop chain configurations, respectively. As shown in this figure, four of these sequences ($s8s$, $1s6s1$, $2s4s2$ and $3s2s3$) can form loops due to intramolecular bonds between the stickers, while the fifth one ($4ss4$) can not.   

\begin{figure*}[h]
\centering
  \includegraphics[width=\linewidth]{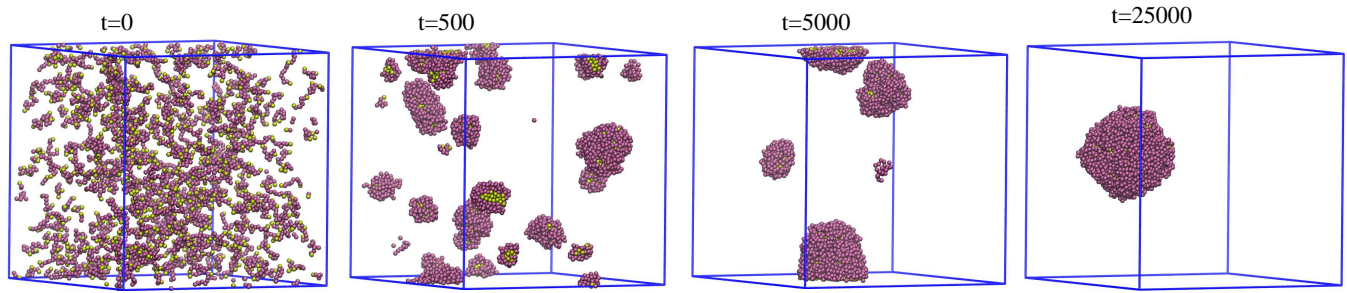}
  \caption{Snapshots of the time evolution from the state of preparation till the formation of a single large droplet in the $s8s$ system. Stickers and non-stickers are represented by yellow and purple spheres, respectively.}
  \label{fgr:drop}
\end{figure*}

We proceed to examine droplet formation in these associating polymer systems. After preparing a random initial state in good solvent, we increased the interaction parameter between the two stickers by a factor of five to $\epsilon_{s}=4.0$. The interaction parameter between the non-sticker beads (and that between stickers and non-stickers) remained $\epsilon_{ns}=0.8$ but all the cutoff distances were increased to $r^{cut}_{ij}=2.5\sigma$. As we have shown before, this choice of interaction parameters guarantees phase separation via formation of polymer droplets. We monitored the evolution of the five systems corresponding to the different sequences  shown in Fig. \ref{fgr:schematic}. Snapshots of one such system (the $s8s$ sequence), from the state of preparation at $t=0$ till $t=25,000$ (in units of LJ time $\tau_{LJ}$), are shown in Fig. \ref{fgr:drop}.  Growth occurs by coalescence of small droplets which are formed by aggregation of neighboring polymers immediately upon quenching the system to poor solvent conditions. This process continues until a single large droplet remains. All other sequences undergo a similar evolution process of droplet formation and growth through coalescence (not shown). 

\begin{figure}[h]
  \includegraphics[width=0.6\linewidth]{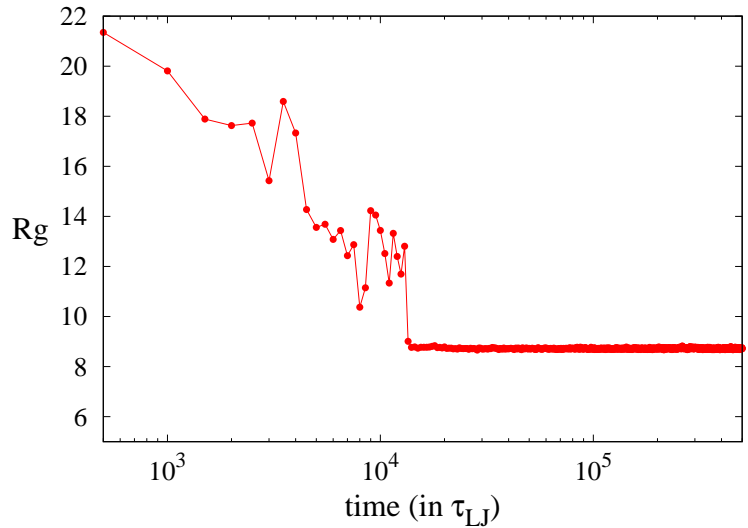}
	\caption{\label{fgr:Rg} Plot of time evolution of radius of gyration of all monomers in the system for the s8s sequence.}
\end{figure}

How long does it take for the final droplet to form? In order to answer this question we monitored the time evolution of the radius of gyration $R_g=\sum_{(i,j)} (r_i-r_j)^2/2N^2$ of all the monomers in the $s8s$ system (see Fig. \ref{fgr:Rg}). Initially, all the polymers are uniformly distributed in the entire simulation box which gives a large value of $R_g$ but as time progresses $R_g$ decreases and eventually saturates at a plateau value which corresponds to the formation of a large droplet that contains all the polymers in the system. The decrease in the $R_g$ value is non-monotonic with time as the system evolves. This happens because of the presence of many droplets during intermediate times (see snapshots at $t=500$ and $t=5,000$ in Fig. \ref{fgr:drop}). The  continuous random motion of these droplets leads to fluctuations of inter-droplet distances and to non-monotonic dependence of $R_g$  on time before it saturates, as shown in figure \ref{fgr:Rg}. Similar time evolution is observed in all other systems with different polymer sequences and in all cases the time it takes a single droplet to form is below $20,000$.    

\begin{figure*}[h]
\centering
  \includegraphics[width=\linewidth]{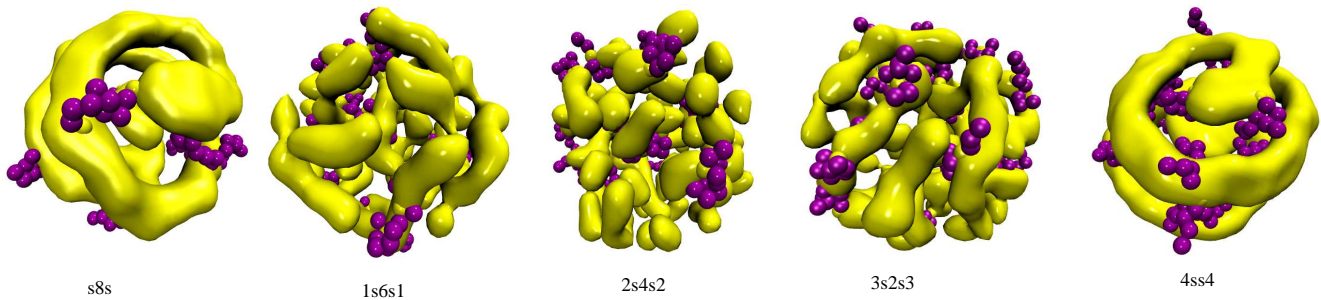}
  \caption{  \label{fgr:fiber} Morphology of structures formed by stickers inside the droplet at $t=300,000$ (shown in green). Non-sticker beads of several polymers are shown as purple spheres.}
\end{figure*}

Having explored the dynamics of droplet formation we proceed to study its local structure in order to characterize the clustering of stickers and the competition between intra and inter-molecular associations of the polymers inside the droplets. To this end, droplets formed at some time $\le 20,000$ are further evolved till $t=300,000$ (till $t=500,000$ for the $s8s$ system) to ensure equilibration. Clusters with very different structures were observed for different sequences (see Fig. \ref{fgr:fiber}). Clusters of stickers were defined operationally as follows: a sticker is assumed to belong to a cluster if it is found within a range of $1.5\sigma$ from any other sticker that belongs to this cluster.

\begin{figure}[h]
  \includegraphics[width=0.5\linewidth]{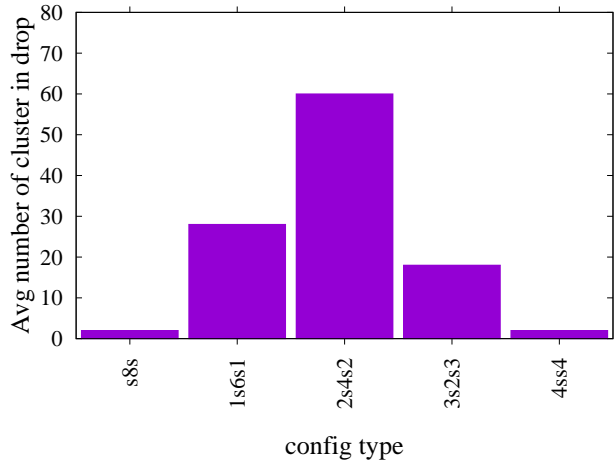}
  \caption{\label{fgr:histo} Histogram showing average number of cluster for all five different sequences at $t=300,000$.}
\end{figure}

Inspection of the $s8s$ droplet shows that almost all stickers belong to a single cluster that has the shape of a long cylindrical fiber which forms a spiral inside the droplet. A more thorough examination revealed the presence of another (smaller) compact cluster at the center of the droplet. A similar spiral fiber formed by the stickers is observed in the $4ss4$ droplet.  In the other three cases corresponding to $1s6s1$, $2s4s2$, and $3s2s3$ sequences, many small elongated clusters whose size and number depends on the sequence, are present in the equilibrium droplet (see figure \ref{fgr:fiber}). For example, the $2s4s2$ sequence has smaller and more numerous clusters compared to sequences $1s6s1$ and $3s2s3$.  Figure \ref{fgr:histo} shows a histogram of the  number of clusters in a droplet for the five different sequences, at time $t=300,000$. In order to test the dependence of our results on polymer concentration, we performed simulations for polymer volume fraction $\phi = 0.006$, for $s8s$ and $2s4s2$ sequences, and did not observe any qualitative changes of size and shape of clusters compared to the $\phi = 0.011$ case shown in Fig. \ref{fgr:fiber}.  
 
Next, we studied the aging of a droplet starting from its formation and following its evolution on timescales that are several orders of magnitude larger than droplet formation time, till it settles into equilibrium.  Two types of structural rearrangements were monitored: on a mesoscopic scale, we followed the association and dissociation of clusters of stickers until a steady state was reached in which the average number and size of these clusters do not change. For all polymer sequences we found that the system attains equilibrium with respect to cluster number and size shortly after the formation of the droplet (not shown). On a molecular level we followed the change of the distribution of polymer states (open and bound states in fig. \ref{fgr:schematic}) as time progresses. To this end we computed the average (over all polymers in the droplet) distance between two stickers of the same polymer at different times. 

\begin{figure*}[h]
\centering
  \includegraphics[width=0.6\linewidth]{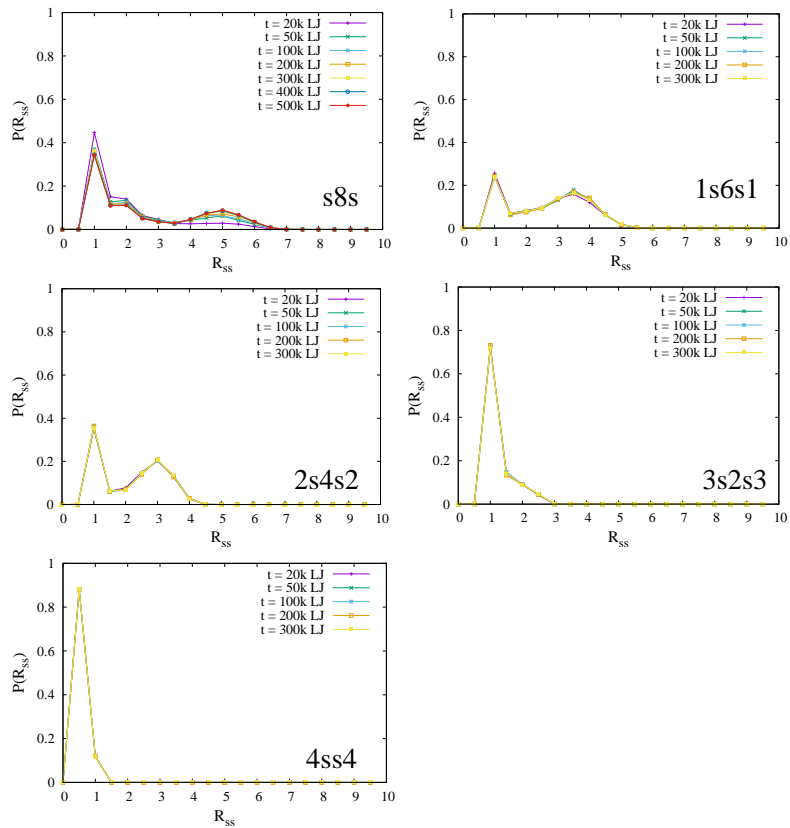}
  \caption{  \label{fgr:Re} Distribution of rms distance $R_{ss}$ between two stickers on a polymer inside the droplet, at different aging times, for five different sequences.}
\end{figure*}

The distributions of ensemble averaged rms distance $R_{ss}$  between two stickers on a polymer, measured at different aging times, are shown in Fig. \ref{fgr:Re} where each of the five panels represents a different sequence. At all aging times the distributions are bimodal for polymers with sequences $s8s$, $1s6s1$ and $2s4s2$ and unimodal for sequence $4ss4$. The $R_{ss}$ distribution for the $3s2s3$ sequence has a single peak followed by a broad shoulder and is intermediate between the bimodal and the unimodal cases. In the bimodal case the two peaks can be associated with open (large $R_{ss}$) and with bound (small $R_{ss}$) states of the corresponding polymer, with the latter peak located at $R_{ss}\approx1$, in agreement with expectations. The position of the second peak that corresponds to open chain configurations (state 1 in Fig.  \ref{fgr:schematic}) is sequence-dependent and increases with the length $n$ of the sequence of beads between the two stickers. In order to understand the origin of this sequence dependence note that the spatial separation between the stickers averaged over open chain conformations, increases with the length $n$ of the sequence of beads between the two stickers (for long sequences and no interaction between the beads one expects the Gaussian chain result, $R_{ss}\propto n^{1/2}$) and therefore, the value of $R_{ss}$ should monotonically decrease from $s8s$ ($n=8$) to $1s6s1$ ($n=6$) to $2s4s2$ ($n=4$) sequence, as observed in Fig. \ref{fgr:Re}.  As expected, only open configurations are observed for $4ss4$ polymers which cannot form intramolecular loops, with a peak at $R_{ss}\approx 0.5$ (the smaller value of $R_{ss}$ is due to attractive FENE bond between neighboring stickers). A single smeared  peak is observed for the $3s2s3$  sequence because of the very small difference between the sticker-sticker distance in open (state 1) and bound (state 2) states. 

\begin{figure}[h]
  \includegraphics[width=0.5\linewidth]{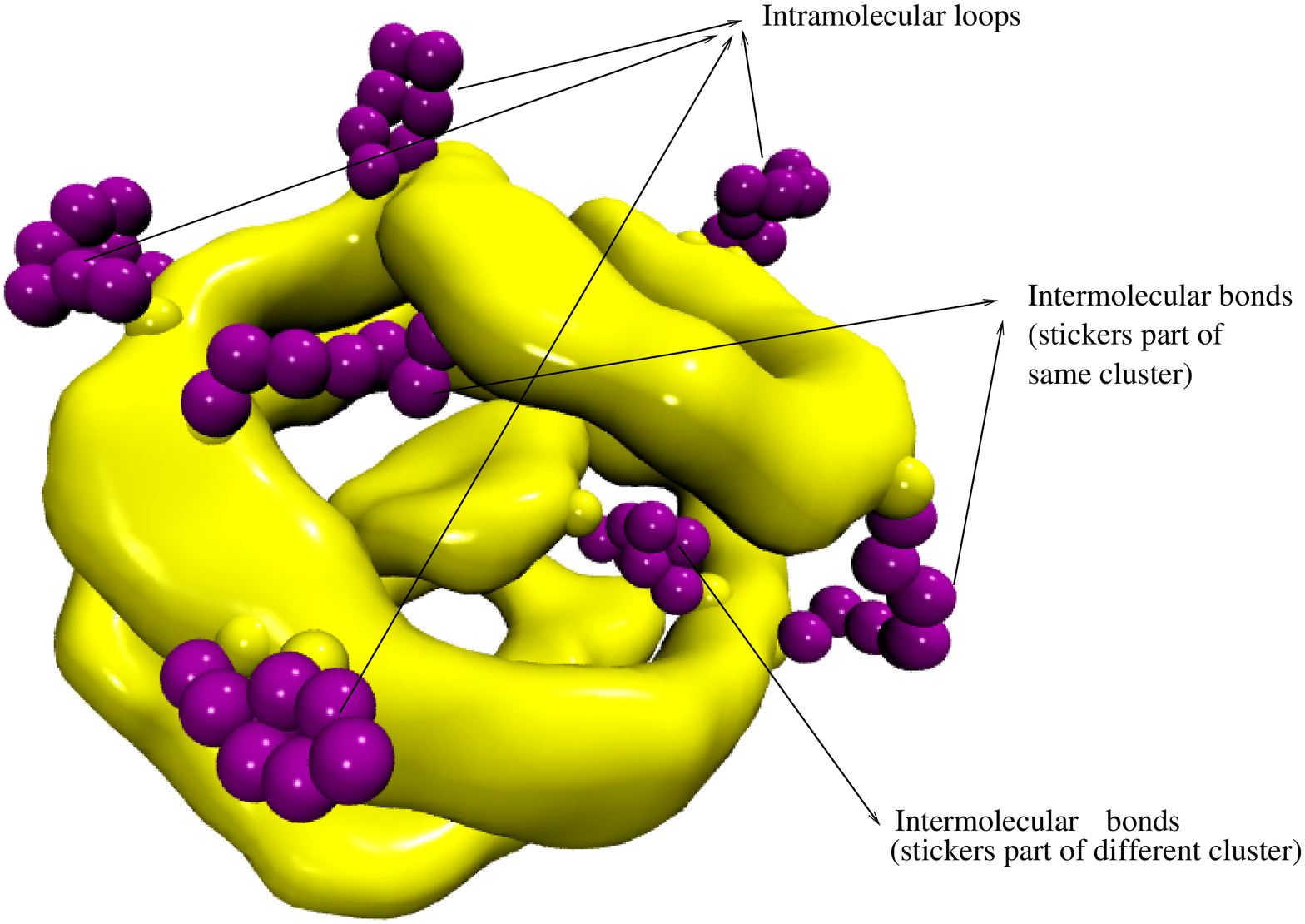}
	\caption{\label{fgr:bridges} A snapshot of clusters of stickers for the $s8s$ sequence. Several intramolecular loops and intermolecular bridges are shown.}
\end{figure}

Note that practically all the polymers are confined within the droplets and all their stickers are arranged in clusters. Therefore, the presence of loops and open chain configurations that correspond to the two peaks (for sequences $s8s$, $1s6s1$ and $2s4s2$) means that some of the polymers form intramolecular loops that bind together to a cluster while other polymers form intermolecular bridges between points on the same or on different clusters (see Fig.   \ref{fgr:bridges}). The numbers of intramolecular loops and intermolecular bridges depends on the sequence (compare the areas under the two peaks in the first three panels in Fig. \ref{fgr:Re}).

We now proceed to examine aging effects. Inspection of Fig. \ref{fgr:Re} shows that for four of the sequences, no significant molecular rearrangement  is observed in the time interval $20,000-300,000$. The only exception is the $s8s$ sequence for which the number of intermolecular bridges increases monotonically with time at the expense of intramolecular loops, and eventually saturates around $t=400,000$.
Even though polymers with $4ss4$ sequence form a single cylindrical fiber that is quite similar to that of  $s8s$ sequence, no aging is observed for $4ss4$ sequence, presumably because it cannot form intermolecular bridges. Clearly, competition between intramolecular loops and intermolecular bridges is a necessary but not a sufficient condition for aging since it is not observed in $1s6s1$ and $2s4s2$ droplets where both open and closed polymer conformations (and multiple small clusters of stickers) are present. 

\begin{figure}[h]
  \includegraphics[width=\linewidth]{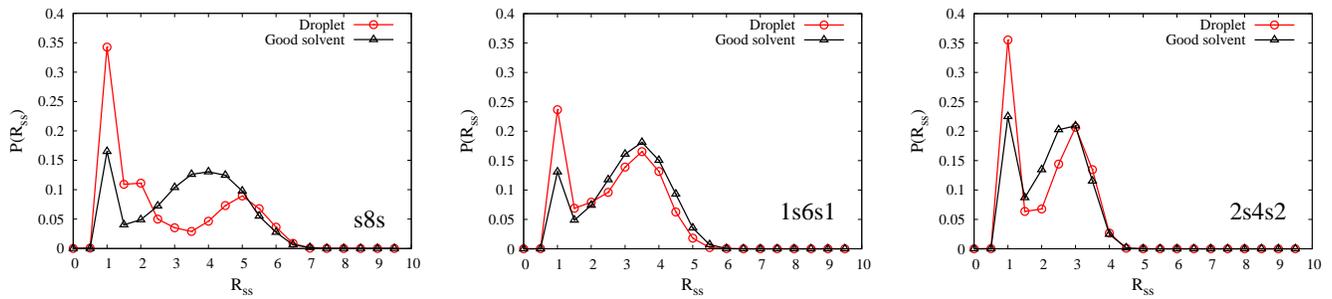}
	\caption{\label{fgr:Rss_compare} Comparison of $R_{ss}$ distribution of polymer chains inside droplets with that of isolated associating polymers in good solvent,  for $s8s$, $1s6s1$ and $2s4s2$ sequences.}
\end{figure}

In order to get some intuition about the effects of intermolecular interactions on the conformation of associating polymers in the droplet, we proceed to compare the $R_{ss}$ distribution of polymers inside the droplet to that of an isolated associative polymer in poor and in good solvent. We first simulated an isolated associative polymer in poor solvent, keeping all parameters the same as in the droplet case discussed above. For four of the five sequences we observed a strong single peak at $R_{ss}=1$ (at $R_{ss}=0.5$ for the $4s4$ sequence), with an extended shoulder for larger $R_{ss}$ values (see figure S1 in SI). This concurs with the expectation that open chain configurations are strongly suppressed and collapsed ones are enhanced for isolated polymers in poor solvent.  
Next, we simulated an isolated associative polymer in good solvent, with $\epsilon_{s}=4.0$ interaction between stickers and only repulsive interactions between the other beads ($\epsilon_{ns}=0.8$ with a cutoff at $r^{cut}_{ij}=2^{1/6}\sigma$). The results are shown in figure \ref{fgr:Rss_compare} where  the $R_{ss}$ distribution of a polymer in a droplet is compared with that in good solvent, for $s8s$, $1s6s1$ and $2s4s2$ sequences. In all these cases a bimodal distribution is observed and the probability of bound state formation is higher in a droplet than in good solvent. For the $1s6s1$ and $2s4s2$ sequences, the open chain peaks are quite similar in amplitude and position in the droplet and in good solvent, but for the $s8s$ sequence the droplet peak is lower and is shifted to higher inter-sticker distances compared to good solvent. A tentative explanation is that in the $1s6s1$ and $2s4s2$ cases there are multiple small clusters of stickers inside the droplet and intermolecular bridges can form between clusters whose separation happens to coincide with the average end-to-end distance of a free polymer chain. In the $s8s$ case the stickers form a long helical fiber that is folded inside the droplet and chains have to stretch in order to form intermolecular bridges between neighboring turns of the fiber. 
Since at $t=0$ we begin with a dilute solution of associating polymers in poor solvent in which most of the chains contain  intramolecular bonds between their stickers, the observation of a second peak that corresponds to intermolecular bridges means that major molecular rearrangement takes place inside droplets formed by polymers with $s8s$, $1s6s1$ and $2s4s2$ sequences. The fact that significant aging after the formation of a large droplet is observed only for the $s8s$ sequence implies that for the $1s6s1$ and $2s4s2$ sequences, the distance between the stickers is optimized by partial conversion from intramolecular to intermolecular associations, already during the growth of smaller droplets and is completed by the time the large droplet is formed by their coalescence.

\begin{figure}[h]
  \includegraphics[width=0.4\linewidth]{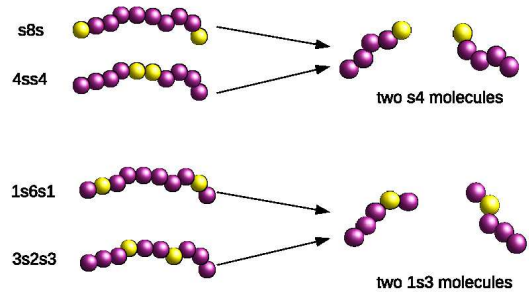}
	\caption{\label{fgr:schematic1} Correspondence between $N=10$ polymers with two stickers and their one-sticker $N=5$ repeat units. }
\end{figure}

Inspection of Figs. \ref{fgr:fiber} and \ref{fgr:histo} shows remarkable similarity between internal droplet morphologies (size, shape and number of clusters of stickers) of sequences $s8s$ and $4ss4$ and of sequences $1s6s1$ and $3s2s3$. In order to understand the origin of this sequence-morphology correlation we splitted each of the $10$-long polymer sequences in the middle to yield two identical sequences of length five, with one sticker each  (see figure \ref{fgr:schematic1}). Clearly, both $s8s$ and $4ss4$ consist of two identical $s4$ repeat units and both $1s6s1$ and $3s2s3$ contain two $1s3$ repeat units. In order to check whether a $N=5$ polymer with a single sticker produces similar internal droplet morphology to the corresponding $N=10$ polymer with two stickers, we performed simulations of droplet formation and aging in dilute solutions of  $N=5$ polymers,  keeping the volume fraction of polymers and other simulation parameters the same as in the $N=10$ case. In accord with expectations, we obtained droplets with similar internal morphology to those of the corresponding $N=10$ sequence (see figure S2 in SI). We therefore conclude that the number, size and shape of clusters is controlled mainly by the sequence of the repeat unit (``monomer'') of the associating polymer (e.g., $s4$) and not by the way these repeat units are joined together to form a polymer (i.e., $s8s$ or $4ss4$). Note, however, that even though internal structure of droplets is similar for corresponding $N=5$ and $N=10$ sequences, there are profound differences between the two cases at the molecular level. While the formation of intramolecular loops and intermolecular bridges is strictly forbidden in droplets formed by single-sticker $N=5$ polymers (e.g., $s4$ sequence), it can be either forbidden or allowed in clusters formed by the corresponding $N=10$ chains, depending on the way they are joined together (forbidden for $4ss4$ and allowed for $s8s$).

\section{Discussion}
In this work, we modeled  intrinsically disordered proteins (IDPs) as short associative polymers with two stickers and studied the effect of the primary sequence of IDPs on their organization inside condensed droplets. We considered all 5 possible sequences of $N=10$ long polymer with two stickers symmetrically positioned along its contour and observed that growth of condensed droplets occurs via coalescence until a single large spherical droplet is formed, in all the systems. We found a striking dependence of the morphology of clusters of stickers on the sequence: while a long helical fiber was observed for both $s8s$ and $4ss4$ sequences, many small compact (and somewhat elongated) clusters were seen for the other three sequences. We also found that the morphology of the clusters is determined mostly by the repeat unit of the associating polymer: both $s8s$ and $4ss4$ ($1s6s1$ and $3s2s3$) sequences are formed from $s4$  repeat unit  and both have a similar internal morphology (a similar conclusion applies to $1s6s1$ and $3s2s3$ sequences that have a common $1s3$ repeat unit). 

For three of the sequences ($s8s$, $1s6s1$ and $2s4s2$) we found that the average spatial distance $R_{ss}$ between the two stickers of a polymer inside the condensed droplet has a bimodal distribution, such that one of the peaks corresponds to intramolecular bonds and the other to intermolecular bridges between clusters (or between different parts of a long fiber of stickers). Only a single peak that corresponds to intramolecular associations between stickers was observed for the other two sequences, $3s2s3$ and $4ss4$, in accord with the expectation that intermolecular bridges can form only for sequences that have sufficiently long spacers between stickers.

Telechelic polymers (but no other sequences in our study) exhibited slow evolution of the distribution of inter-sticker distances inside the droplet, as the peak corresponding to small (large) values of $R_{ss}$ decreased (increased) with time. We identified this ``aging''  phenomenon with molecular rearrangement  in which loops formed by intramolecular association open up to form intermolecular bridges between different folds of the helical fiber of stickers inside the droplet. The evolution from intramolecular to intermolecular associations in droplets of telechelic polymers in poor solvent should be contrasted with the opposite phenomenon that was observed in gels produced by self assembly of telechelic polymers in good solvent \cite{Kulveer2020}.

Finally we would like to comment on the relevance of our results to biomolecular condensates of intrinsically disordered proteins. Clearly our simple model misses many of the molecular details of real IDPs such as  hydrogen bonding and electrostatic interactions between amino acids and formation of partial secondary structure (such as $\alpha$-helices and $\beta$-strands). Still, it does capture some of the generic feature of flexible proteins such as their ability to change their conformation in order to associate with other macromolecules (in our model this is illustrated by the replacement of intramolecular by intermolecular associations between stickers). It also captures, albeit qualitatively, the phenomenon of aging of liquid-like IDP droplets and in particular their tendency to form strong gels and fibers following a long incubation time. Such phenomena have been observed in liquid droplets of tau proteins \cite{Wegmann2018} and of nucleoporin domains \cite{Gorlich2010, Lemke2020}. Another similarity between our associative polymer model and IDPs is illustrated by the fact that, just like in our simple model, fibril formation by amyloid peptides \cite{Lopez2004}, formation of nanostructures via self-assembly of short peptides \cite{Zhang2002} and nanoscale organization of nucleoporins in the nuclear pore complex \cite{Huang2020}, strongly depend on their sequence. 

\section{acknowledgement}
YR would like to acknowledge helpful discussions and correspondence with Eduard Lemke. This work was supported by grants from the Israel Science Foundation 178/16 and from the Israeli Centers for Research Excellence program of the Planning and Budgeting Committee 1902/12.

\bibliography{references}

\end{document}